\documentclass[journal]{IEEEtran}
\usepackage{pifont}
\usepackage{}
\usepackage[noadjust]{cite}
\usepackage{graphicx}
\usepackage{float}
\usepackage{amsmath}
\usepackage{enumerate}
\usepackage{amssymb}
\usepackage{fixltx2e}
\usepackage{color}
\usepackage{algorithm}
\usepackage{algorithmic}
\usepackage{multirow}
\usepackage{ragged2e}
\usepackage{booktabs}
\usepackage{subfigure}
\usepackage{amsthm}
\usepackage{stfloats}
\usepackage{amsmath,lipsum}

\usepackage{xcolor,cite,etoolbox}
\makeatletter
\pretocmd\@bibitem{\color{black}\csname keycolor#1\endcsname}{}{\fail}
\newcommand\citecolor[1]{\@namedef{keycolor#1}{\color{blue}}}
\makeatother

\begin{document}
\pagenumbering{arabic}

\title{{\huge{Low Complexity First: Duration-Centric ISI Mitigation in Molecular Communication via Diffusion}}}

\author{Xuan Chen,  Fei Ji,~\emph{Member,~IEEE,} Miaowen Wen,~\emph{Senior Member,~IEEE,}\\
Yu Huang, Yuankun Tang,~Andrew W. Eckford,~\emph{Senior Member,~IEEE} \\

\thanks{X. Chen is with the School of Electronics and Information Engineering,
South China University of Technology, Guangzhou 510641, China, and also with the Department of Electrical Engineering and Computer Science, York University, Toronto M3J 1P3, Canada (Email:eechenxuan@mail.scut.edu.cn).

    F. Ji, M. Wen, and Y. Tang are with the School of Electronic and Information Engineering, South China University of Technology, Guangzhou 510640, China (Email: \{eefeiji, eemwwen\}@scut.edu.cn, eeyktang@mail.scut.edu.cn).

    Y. Huang is with the School of Electronics and Communication Engineering, Guangzhou University, Guangzhou 510006, China
    (Email: yuhuang@gzhu.edu.cn).

    A. W. Eckford is with the Department of Electrical Engineering and Computer Science, York University, Toronto M3J 1P3, Canada (Email: aeckford@yorku.ca).
}}

\maketitle
\vspace{-0.3cm}
\begin{abstract}

 In this paper, we propose a novel inter-symbol interference (ISI) mitigation scheme for molecular communication via diffusion (MCvD) systems with optimal detection interval. Its rationale is to exploit the discarded duration (i.e., the symbol duration outside this optimal interval) to relieve ISI in the target system. Following this idea, we formulate an objective function to quantify the impact of the discarded time on bit error rate (BER) performance. Besides, an optimally reusable interval within the discarded duration is derived in closed form, which applies to both the absorbing and passive receivers. Finally, numerical results validate our analysis and show that for the considered MCvD system, significant BER improvements can be achieved by using the derived reusable duration.

\end{abstract}


\date{\today}
\renewcommand{\baselinestretch}{1.2}
\setcounter{page}{1}

\IEEEpeerreviewmaketitle

\section{Introduction}

As a new communication paradigm proposed in recent years, molecular communication via diffusion (MCvD) exhibits a range of advantages over traditional communication methods, such as, small size, high energy efficiency, and excellent bio-compatibility \cite{MC}. MCvD is expected to serve certain scenarios where other communication schemes are inappropriate or unusable, such as the nano-scale communication and biomedical fields \cite{Reviewer_1}. However, due to the long delay spread of molecular diffusion in a fluid medium, MCvD systems suffer from severe inter-symbol interference (ISI).

To alleviate this defect, various ISI cancellation and suppression schemes have been proposed and studied, which can be categorized in three ways: modulation-based \cite{modulation_based_1,modulation_based_2}; equalization-based \cite{equalization_based_3}; and channel-based, i.e., introducing an external factor, such as flow~\cite{Flow_ISI_mitigation_1}~or~enzyme~\cite{Enzyme_ISI_mitigation_1}, into the channel. Besides the above schemes, ISI can also be mitigated by adjusting the detection interval, such as by the shift-$\tau$ method \cite{The_shift_method_1}; truncating the symbol duration in advance \cite{Truncation_1}; and extracting a small portion of the symbol duration \cite{Extracting_1}. Compared with the ISI mitigation schemes in \cite{modulation_based_1,modulation_based_2,equalization_based_3,Flow_ISI_mitigation_1,Enzyme_ISI_mitigation_1}, the proposed schemes in \cite{The_shift_method_1,Truncation_1,Extracting_1} are more appropriate for MCvD systems, given the computational constraints of tiny nano-machines.

In our own previous work, we optimized the detection interval and derived a closed-form solution for this optimal interval in \cite{Xuan_previous_work}. Simulation results demonstrated that MCvD systems with the optimal detection interval have a competitive advantage in terms of bit error rate (BER). This optimal interval is determined based on the criterion that the desired signal dominates the ISI, which suggests that it is usually shorter than a symbol duration.
However, a question naturally arises: whether there exist features in the discarded duration that are useful to signal recovery. In particular, the discarded duration represents the symbol interval other than the optimal detection interval, in which the ISI dominates the signal. Therefore, it has the potential to be used to estimate and mitigate ISI in the optimal detection interval.

In this paper, we address this question directly by exploring the feasibility of reusing the discarded time to aid signal recovery. Specifically, we use the received signal from the discarded time to counteract some of the ISI in the optimal detection interval. Further, we construct a objective function to describe the impact of the discarded time on BER performance. Accordingly, an optimally reusable duration within this discarded interval is derived in closed form for all considered receivers. Monte Carlo simulations are performed to verify the theoretical analysis and compare the performance of MCvD systems with/without the reusable duration, where the optimal detection interval is always applied.


\section{System Model and Problem Statement}\label{System Model and Problem Statement}


\subsection{System Model}
In this paper, we consider a typical MCvD system consisting of a point transmitter and a spherical receiver, where the receiver can be fully absorbing or passive. We assume that the transceiver is placed in an unbounded 3-dimensional (3D) environment, in which the distance from the transmitter to the closest point of the receiver's surface is $d$ and the receiver's radius is $r$. It is also assumed that the on-off keying (OOK) is applied, where the transmitter releases $Q$ molecules to convey symbol ``1'', while releasing no molecules to convey symbol ``0''. Besides, we assume that perfect time synchronization can be achieved. In the following, we review the preliminary conceptual framework for MCvD systems with different receivers.


\subsubsection{Fundamentals of Absorbing Receiver}

Following Fick's law of diffusion, the probability of a molecule, released from a point transmitter at $t = 0$, reaching a spherical receiver at time $t$ is
\newpage
\begin{align}\label{CIR_absorbing}
h(t)=\frac{r}{d+r} \frac{d}{\sqrt{4 \pi D t^{3}}} \exp \left(-\frac{d^{2}}{4 D t}\right),
\end{align}
where $D$ is the diffusion coefficient of information molecules.
Then we can express the expected fraction of molecules, absorbed by the receiver in $[t_1,t_2]$ with $t_1, t_2 \in [0, T_s]$ and $t_2 > t_1$, as
\begin{align}\label{CIR_absorbing_for_interval}
\hspace{-0.2cm}F{\left(t_{1}, t_{2}\right)}=\frac{r}{d+r}\left[\operatorname{erf}\left(\frac{d}{\sqrt{4 D t_{1}}}\right)-\operatorname{erf}\left(\frac{d}{\sqrt{4 D t_{2}}}\right)\right],
\end{align}
where $T_s$ denotes the symbol duration and ${\rm{erf}}\left(  \cdot  \right)$ is the error function. Let us define $x_k$ and $Y_k$ as the $k$-th transmitted bit and the number of received molecules corresponding to the $k$-th transmission, respectively, where $k=1,2,\cdots$. Then we approximately have
\begin{align}\label{Approximated_Absorbing_molecules_1}
Y_k^{ab}{\rm{ = }}\underbrace {Q{x_{k - i}}F_{\left( {{t_1},{t_2}} \right)}^i}_{{\rm{desired~signal}}}{\rm{ + }}\underbrace {\sum\limits_{i = {\rm{1}}}^{\min \left\{ {L,k} \right\}} Q {x_{k - i}}F_{\left( {{t_1},{t_2}} \right)}^i}_{{\rm{ISI~signal}}}{\rm{ + }}\underbrace {n_k^{ab}}_{\rm{noise}}
\end{align}
where $F_{\left( {{t_1},{t_2}} \right)}^i = F\left( {{t_1} + i{T_s},{t_2} + i{T_s}} \right)$, $L$ is the ISI length, and ${n_k^{ab}}$ is assumed to follow the Gaussian distribution, i.e.,
\begin{align}
    n_k^{ab}\sim \sum\limits_{i = 0}^{\min \left\{ {L,k} \right\}} Q {x_{k - i}}{\cal N}\big( {0,\sigma _{\left( {{t_1},{t_2}} \right)}^i} \big)
\end{align}
with $\sigma _{\left( {{t_1},{t_2}} \right)}^i{\rm{ = }}F_{\left( {{t_1},{t_2}} \right)}^i\big( {1 - F_{\left( {{t_1},{t_2}} \right)}^i} \big)$. For clarity, \eqref{Approximated_Absorbing_molecules_1} can be rewritten as
\begin{align}\label{Approximated_Absorbing_molecules}
Y_k^{ab}\sim\sum\limits_{i = 0}^{\min \left\{ {L,k} \right\}} Q {x_{k - i}}{\cal N}\left( {F_{\left( {{t_1},{t_2}} \right)}^i,\sigma _{\left( {{t_1},{t_2}} \right)}^i} \right).
\end{align}
Besides, we assume that the energy detection is used for the absorbing receiver and then the average bit error probability $P_e$ can be written as [11, eq.~5].

\subsubsection{Fundamentals of Passive Receiver}

Similar to the absorbing receiver, we can define when $\frac{r}{r+d} <0.15$, the probability of observing a given molecule, emitted from the point transmitter at $t = 0$, inside $V$ at time $t$ as \cite{passive_receiver_probability}
\begin{align}\label{Passive_CIR}
p\left( t \right) = \frac{V}{{{{\left( {4\pi Dt} \right)}^{3/2}}}}\exp \left( { - \frac{{{{\left( {d + r} \right)}^2}}}{{4Dt}}} \right),
\end{align}
where $V$ is the volume of the passive receiver. First, we assume that $N$ samples are taken by the receiver at a symbol duration given by $f(n) \in \left[ {0,{T_s}} \right]$ where $n = 0,1, 2,\cdots,N$, and that they are equally summed up before the single threshold detection. The number of received molecules for the $k$-th transmission can be approximately expressed as
\begin{align}\label{Approximated_Counting_molecules}
{Y_k^{pa}} \sim \sum\limits_{i = 0}^{\min \left\{ {L,k} \right\}} {\sum\limits_{n = {n_1}}^{{n_2}} {{Q}{x_{k - i}}{\mathcal N}\left( {{p_{n,i}},{p_{n,i}}} \right)} } ,
\end{align}
where $p_{n,i} = p\left( {f\left( n \right) + i{T_s}} \right)$; $n_1$ and $n_2$ denote the first sample and the last sample employed for the considered MCvD system, respectively, with $0\le{n_1},{n_2}\le N$. Besides, we assume $f(n) = nt_s$ and $t_s = T_s/N$, where $n \in \left[ {{n_1},{n_2}} \right]$. The average bit error probability $P_e$ can also be found in \cite{Xuan_previous_work}.

\subsection{Problem Statement}


\begin{figure}[t]
	\centering
	 \includegraphics[width=3.7in]{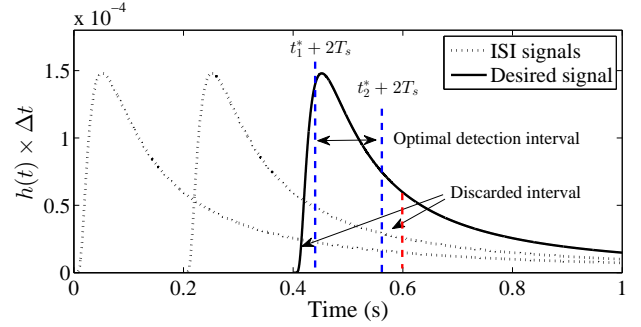}
	\caption{Desired and undesired CIRs for the absorbing receiver with the optimal detection interval, where $T_s = 0.2$, $\Delta t = 10^{-4}$, and other system parameters are listed in Table I of \cite{Xuan_previous_work}.}
	\label{figure-CIR_AB}
\end{figure}

According to the previous research, the BER can be significantly reduced by optimizing the detection interval for MCvD systems. Yet, whether the received signal outside of this interval is necessarily destructive to signal recovery? To answer this question, we take the absorbing receiver as an example to explore the feasibility of reusing the discarded time to assist the detection.

For clarity, we plot the desired and undesired channel impulse responses (CIRs) obtained from \eqref{CIR_absorbing} in Fig.~\ref{figure-CIR_AB}, where the optimal detection interval and the discarded interval are defined as $\left[ {t_1^*,t_2^*} \right]$ and $\left[ {0,t_1^*} \right) \cup \left( {t_2^*,{T_s}} \right]$, respectively, for the current transmission. Clearly, the ISI signals play a leading role in the received signal when $t \in \left[ {0,t_1^*} \right) \cup \left( {t_2^*,{T_s}} \right]$. This enlightens us on whether the received signal from $\left[ {0,t_1^*} \right) \cup \left( {t_2^*,{T_s}} \right]$ can be used to counteract some of the ISI in ${\left[ {{t_1^*},{t_2^*}} \right]}$, since the ISI symbols $\left[ {{x_{k - {\min \left\{ {L,k} \right\}}}}, \cdots, {x_{k - 2}},{x_{k - 1}}} \right]$ remain unchanged during the $k$-th transmission. If feasible, we will perform a subtraction between the received signals in ${\left[ {{t_1^*},{t_2^*}} \right]}$ and $T_r$, where $T_r$ denotes a potentially reusable duration within the discarded interval, i.e., ${T_r} \subset \left[ {0,t_1^*} \right) \cup \left( {t_2^*,{T_s}} \right]$. Then, from \eqref{Approximated_Absorbing_molecules}, we have
\begin{align}\label{Updated_Approximated_Absorbing_molecules}
\hat Y_k^{ab}\sim{\hspace{-0.2cm}}\sum\limits_{i = 0}^{\min \left\{ {L,k} \right\}} Q {x_{k - i}}{\cal N}\left( {F_{\left( {t_1^*,t_2^*} \right)}^i{\rm{ - }}F_{{T_r}}^i,\sigma _{\left( {t_1^*,t_2^*} \right)}^i{\rm{ + }}\sigma _{{T_r}}^i} \right).
\end{align}
Comparing \eqref{Approximated_Absorbing_molecules} and \eqref{Updated_Approximated_Absorbing_molecules}, we can find that the ISI in $\left[ {t_1^*,t_2^*} \right]$ has dropped by $\sum\nolimits_{i = {\rm{1}}}^{\min \left\{ {L,k} \right\}} {Q{x_{k - i}}F_{{T_r}}^i} $ due to $F_{\left( { \cdot , \cdot } \right)}^i \ge 0$. However, it is worth noting that the desired signal and noise in \eqref{Updated_Approximated_Absorbing_molecules} also change. Hence, how to choose a proper $T_r$ still has a long way to go.

First, let us assume ${T_r} \subset \left({t_2^*,{T_s}} \right]$. After the subtraction in \eqref{Updated_Approximated_Absorbing_molecules}, we can observe that the desired signal will be highly attenuated in $\hat Y_k^{ab}$, since the expected signal belonging to $T_r$ is still relatively strong. Consequently, $\left({t_2^*,{T_s}} \right]$ cannot be viewed as a properly reusable duration. Next, we consider ${T_r} \subset \left[ {0, t_1^*} \right)$. When $t \in \left[ {0,t_1^*} \right)$, it can be seen from Fig.~\ref{figure-CIR_AB} that there is a short period in which the ISI signal is predominant in all received signals. This means that it is possible to use the signal received in the mentioned time to mitigate the ISI in $\left[ {t_1^*,t_2^*} \right]$ while preserving the advantage of desired signals. For ease of analysis, we refer to the above period as $\left[ {0,{t_u}} \right]$, i.e., ${T_r}{\rm{ = }}\left[ {0,t_u} \right]$, and $t_u$ is defined as the cut-off value of the reusable duration with ${t_u} \in \left[ {0,t_1^*} \right)$. The next step is to find an optimal $t_u$ from all possible values, formulated as
\begin{align}\label{objective_AB}
{t_u^*}=\hspace{-0.1cm} \mathop {\arg \min }\limits_{0 \le t_u < t_1^*} {P_e} .
\end{align}
We can observe from [11, eq.~5] that it is challenging to obtain ${t_u^*}$ without the aid of the exhaustive search due to the complex expression of $P_e$. Based on the study in \cite{Xuan_previous_work}, we propose to use a performance metric, namely modified signal-to-interference and noise amplitude ratio (mSINAR), to simplify the solving process of $t_u^*$ in the sequel. mSINAR can be defined as
\begin{align}\label{mSINAR}
{\textrm{mSINAR}} = \frac{{\frac{1}{2}F_{\left( {{t_1},{t_2}} \right)}^0}}{{\sum\limits_{k = 1}^L {\frac{1}{2}} F_{\left( {{t_1},{t_2}} \right)}^k + \sum\limits_{k = 0}^L {\sqrt {\frac{{F_{\left( {{t_1},{t_2}} \right)}^k\left( {1 - F_{\left( {{t_1},{t_2}} \right)}^k} \right)}}{{2{Q}}}} } }},
\end{align}
and the value of mSINAR is in the range of $\left( {0,1} \right]$. For clarity, we set a cut-off point $\hat Q$ to divide the valid value of~mSINAR and it can be calculated from \eqref{mSINAR} with mSINAR~$=$~1.
Specifically, mSINAR $\in \left( {0,1} \right)$ corresponds to $0< Q < \hat Q$, while mSINAR $=$ 1 means $Q \ge \hat Q$.

\begin{figure*} [ht]
\begin{align}\label{problem_formulation_based_on_mSINAR}
{\tilde t}_u^* =\left\{
\begin{array}{l}
\mathop {\arg \max }\limits_{0 \le {t_u} < t_1^*} \frac{{F_{\left( {t_1^*,t_2^*} \right)}^0 - F_{\left( {0,t_u} \right)}^0}}{{\sum\limits_{k = 1}^L {\left( {F_{\left( {t_1^*,t_2^*} \right)}^k - F_{\left( {0,t_u} \right)}^k} \right)}  + \sqrt {\frac{2}{{Q}}} \sum\limits_{k = 0}^L {\sqrt {F_{\left( {t_1^*,t_2^*} \right)}^k\left( {1 - F_{\left( {t_1^*,t_2^*} \right)}^k} \right) + F_{\left( {0,t_u} \right)}^k\left( {1 - F_{\left( {0,t_u} \right)}^k} \right)} } }},~{\textrm{if}}~0 < {Q} < \hat{Q} \\
\\
\vspace{-0.66cm}
\\
\mathop {\arg \max }\limits_{0 \le {t_u} < t_1^*} \frac{{F_{\left( {t_1^*,t_2^*} \right)}^0 - F_{\left( {0,t_u} \right)}^0}}{{\sum\limits_{k = 1}^L {\left( {F_{\left( {t_1^*,t_2^*} \right)}^k - F_{\left( {0,t_u} \right)}^k} \right)}  + \sqrt {\frac{2}{{\hat{Q}}}} \sum\limits_{k = 0}^L {\sqrt {F_{\left( {t_1^*,t_2^*} \right)}^k\left( {1 - F_{\left( {t_1^*,t_2^*} \right)}^k} \right) + F_{\left( {0,t_u} \right)}^k\left( {1 - F_{\left( {0,t_u} \right)}^k} \right)} } }},~{\textrm{if}}~{Q} \ge \hat{Q} \\
\end{array} \right.
\end{align}
\hrulefill
\end{figure*}

\section{Optimization Analysis}\label{Optimization Analysis}

In this section, we use the mSINAR-based approximation methods to solve the optimally reusable duration.

\subsection{Reusable Duration Optimization for Absorbing Receiver}

First, we will describe the detailed procedures to calculate $t_u^*$ for the absorbing receiver. Based on the usage strategy of mSINAR in \cite{Xuan_previous_work}, the objective function in \eqref{objective_AB} can be rewritten as \eqref{problem_formulation_based_on_mSINAR}, shown at the top of the next page.
Here, ${\tilde t}_u^*$ is the approximation of ${t}_u^*$. For clarity, the derived $t_u^*$ for all possible $L$ is described below.

\textbf{\emph{Proposition 1:}} Assuming that the approximation for \eqref{problem_formulation_based_on_mSINAR} for all considered $L$ is exact, the solution of \eqref{problem_formulation_based_on_mSINAR} can be written~as
\begin{align}\label{final_tu_AB}
{t}_u^* \approx {\tilde t}_u^* = \min \left[ {\frac{{ - \beta  + \sqrt {{\beta ^2} + 4\alpha } }}{{2\alpha }},\bar t_1^*} \right],
\end{align}
where $\alpha  = \frac{{51{T_s} - 51{T_s}\ln {\cal I} - 15{m^2}{T_s}}}{{14{m^2}T_s^2}}$, $\beta  = \frac{{60{T_s} - 14{T_s}\ln {\cal I} - 37{m^2}}}{{14{m^2}{T_s}}}$, $\hat t_u^* \approx \frac{{ - \left( {60{T_s} - 37{m^2}} \right) + \sqrt {{{\left( {60{T_s} - 37{m^2}} \right)}^2} + 56\left( {51{T_s} - 15{m^2}{T_s}} \right){m^2}{T_s}} }}{{\left( {51{T_s} - 15{m^2}{T_s}} \right)}}$, $m = \frac{d}{{\sqrt {4D} }}$, $\mathcal{I} = \sum\limits_{k = 1}^L {\left\{ {\frac{{h\left( {k{T_s} + \hat t_u^*} \right)}}{{h\left( {{T_s} + \hat t_u^*} \right)}}} \right\}} $,
and $\bar t_1^*$ is the value to which $t_1^*$ eventually converges.

\emph{Proof:} Please see Appendix A.

\subsection{Reusable Duration Optimization for Passive Receiver}

In this subsection, we focus on calculating $n_u^*$ for the passive receiver, where $n_u^*$ is the optimal cut-off value of the reusable sampling. According to the description on the absorbing receiver, the objective function can be written~as
\eqref{problem_formulation_based_on_mSINAR_passive}, where ${\tilde n}_u^*$ is the approximation of ${n}_u^*$ due to the use of mSIANR.
Here, ${\left[ {{n_1^*},{n_2^*}} \right]}$ denotes the optimal detection sampling. For clarity, the derived $n_u^*$ for all possible $L$ is described below.
\begin{figure*}[htp]
\begin{align}\label{problem_formulation_based_on_mSINAR_passive}
{\tilde n}_u^* = \left\{ \begin{array}{l}
 \mathop {\arg \max }\limits_{0 \le {n_u} < n_1^*} \frac{{\sum\limits_{n = n_1^*}^{n_2^*} {{p_{n,0}}}  - \sum\limits_{n = 0}^{n_u^*} {{p_{n,0}}} }}{{\sum\limits_{k = 1}^L {\left( {\sum\limits_{n = n_1^*}^{n_2^*} {{p_{n,k}}}  - \sum\limits_{n = 0}^{n_u^*} {{p_{n,k}}} } \right)}  + \sum\limits_{k = 0}^L {\sqrt {\frac{2}{Q}\left({\sum\limits_{n = n_1^*}^{n_2^*} {{p_{n,k}}}  + \sum\limits_{n = 0}^{n_u^*} {{p_{n,k}}} }\right) } } }},~{\textrm{if}}~0 < {Q} < \hat {Q}\\
  \\
\vspace{-0.99cm}
\\
  \mathop {\arg \max }\limits_{0 \le {n_u} < n_1^*}\frac{{\sum\limits_{n = n_1^*}^{n_2^*} {{p_{n,0}}}  - \sum\limits_{n = 0}^{n_u^*} {{p_{n,0}}} }}{{\sum\limits_{k = 1}^L {\left( {\sum\limits_{n = n_1^*}^{n_2^*} {{p_{n,k}}}  - \sum\limits_{n = 0}^{n_u^*} {{p_{n,k}}} } \right)}  + \sum\limits_{k = 0}^L {\sqrt {\frac{2}{\hat {Q}}\left({\sum\limits_{n = n_1^*}^{n_2^*} {{p_{n,k}}}  + \sum\limits_{n = 0}^{n_u^*} {{p_{n,k}}} } \right)} } }},~{\textrm{if}}~{Q} \ge \hat {Q}\\
 \end{array} \right.
\end{align}
\hrulefill
\end{figure*}

\textbf{\emph{Proposition 2:}} Assuming that the approximation for \eqref{problem_formulation_based_on_mSINAR_passive} for all considered $L$ is exact, the solution of \eqref{problem_formulation_based_on_mSINAR_passive} can be written~as
\begin{align}\label{final_tu_passive}
n_u^* \approx{\tilde n}_u^* = \min \left[ {\left\lfloor {\frac{{ - \hat \beta  + \sqrt {{{\hat \beta }^2} + 4\hat \alpha } }}{{2\hat \alpha {t_s}}}} \right\rfloor ,\bar n_1^* - 1} \right],
\end{align}
where $\hat \alpha  = \frac{{51{T_s} - 51{T_s}\ln {\cal W} - 15{{\hat m}^2}{T_s}}}{{14{{\hat m}^2}T_s^2}}$, $\hat \beta  = \frac{{60{T_s} - 14{T_s}\ln {\cal W} - 37{{\hat m}^2}}}{{14{{\hat m}^2}{T_s}}}$, \\$\hat n_u^* \approx \left\lfloor {\frac{{ - \left( {60{T_s} - 37{{\hat m}^2}} \right) + \sqrt {{{\left( {60{T_s} - 37{{\hat m}^2}} \right)}^2} + 56\left( {51{T_s} - 15{{\hat m}^2}{T_s}} \right){{\hat m}^2}{T_s}} }}{{\left( {51{T_s} - 15{{\hat m}^2}{T_s}} \right){t_s}}}} \right\rfloor$, $\hat m = \frac{{d + r}}{{\sqrt {4D} }}$, ${\cal W} = \sum\limits_{k = 1}^L {{\frac{{{p_{\hat n_u^*,k}}}}{{{p_{\hat n_u^*,1}}}}} } $,
and $\bar n_1^*$ is the value of the final convergence of $n_1^*$.

\emph{Proof:} Please see Appendix B.

\section{Numerical results and analysis}\label{Numerical results and analysis}



In this section, we perform Monte Carlo simulations to evaluate the BER performance of the proposed scheme. The MCvD system with an optimal detection interval investigated in \cite{Xuan_previous_work} is chosen for comparison. Besides, the conventional OOK scheme and the ideal scheme that uses $t_u^*$ or $n_u^*$ obtained from the exhaustive search for $\arg \min {P_e}$ are selected as benchmarks. Note that the optimal detection threshold obtained by the exhaustive search is assumed to be employed for all schemes. Moreover, the system parameters are listed in Table~I of \cite{Xuan_previous_work} and we set the sampling interval as $t_s = \left\lfloor {\frac{t_{\max}}{6}} \right\rfloor$.

\begin{figure}[t]
    \centering
    \subfigure[absorbing receiver with $T_s = 0.2$s]{
        \includegraphics[width=3.65in]{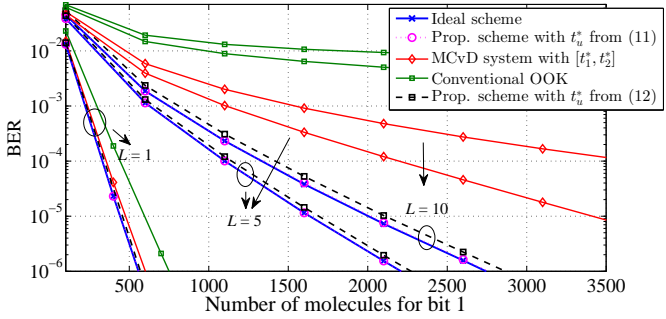}
    \label{Comparison_Ts_0_2}
    }
        \subfigure[absorbing receiver with $T_s = 0.3$s]{
	\includegraphics[width=3.65in]{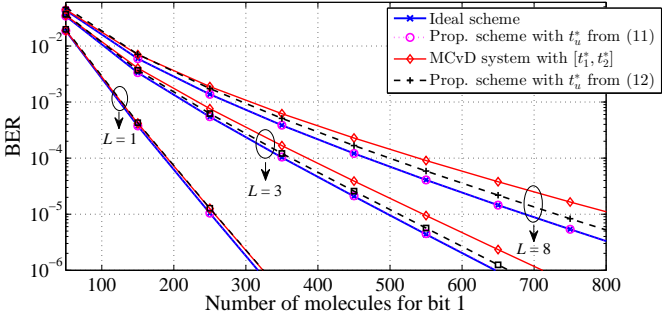}
	\label{Comparison_Ts_0_3}
	}
    \caption{BER performance comparison among different schemes for the absorbing receiver with $T_s = \left\{0.2, 0.3\right\}$s.}
    \label{Absorbing_receiver_comparison}
\end{figure}

Fig.~\ref{Absorbing_receiver_comparison} shows the BER performance comparison among all schemes mentioned previously, where the absorbing receiver with $T_s = \left\{0.2, 0.3\right\}$s is considered. First, we can observe an excellent match between the ideal scheme and the proposed scheme with numerical $t_u^*$ from \eqref{problem_formulation_based_on_mSINAR} for all considered cases, which proves that mSINAR can be a good performance metric to measure the BER performance for the considered MCvD system. As expected, the proposed scheme performs best among all considered schemes. Considering the poor performance of the conventional OOK scheme, its BER curve has been removed for the clarity of the following figures. By observing Fig.~\ref{Comparison_Ts_0_2} and Fig.~\ref{Comparison_Ts_0_3}, one can easily discover that the performance gain achieved by the proposed scheme over other benchmarks is proportional to the ISI length $L$ (inversely proportional to $T_s$).
This is because when the ISI is increasingly serious, the received signal in $\left[ {0,t_u^*} \right]$ contains more interference elements, thus enhancing the capability of mitigating the interference in $\left[ {{t_1^*},{t_2^*}} \right]$. Finally, we can find that the BER curves corresponding to the theoretical $t_u^*$ obtained from \eqref{final_tu_AB} agree with that of the numerical counterparts approximately, verifying the effectiveness of the derived $t_u^*$. Particularly, when $L=1$, these two curves almost coincide; while as $L$ increasingly grows, the gap between these two curves is gradually widening. This is because when solving \eqref{further_simpified_problem_formulation_based_on_mSID}, we use the strongest ISI signal to approximate the remaining ISI signal, which indicates that the increase of $L$ will further weaken the accuracy of the above approximation.

\begin{figure}[t]
    \centering
    \subfigure[passive receiver with $T_s = 1$s]{
        \includegraphics[width=3.63in]{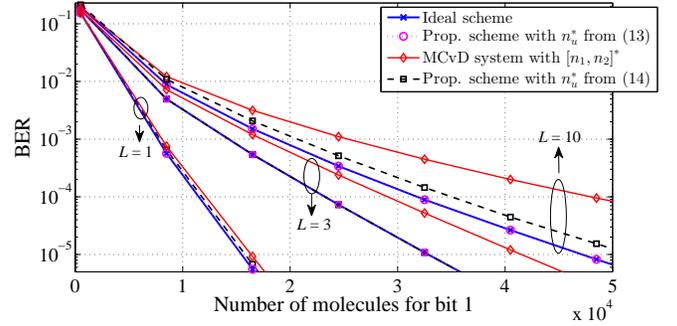}
    \label{Comparison_Ts_1}
    }
        \subfigure[passive receiver with $T_s = 1.5$s]{
	\includegraphics[width=3.63in]{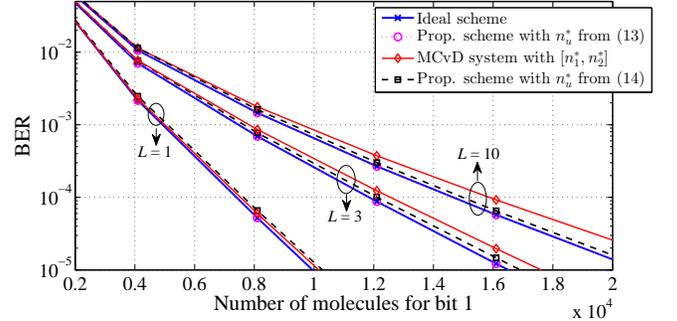}
	\label{Comparison_Ts_1_5}
	}
    \caption{BER performance comparison among different schemes for the passive receiver with $T_s = \left\{1, 1.5\right\}$s.}
    \label{Passive_receiver_comparison}
\end{figure}

Fig.~\ref{Passive_receiver_comparison} gives the comparison results similar to Fig.~\ref{Absorbing_receiver_comparison}, where the passive receiver with $T_s = \left\{1, 1.5\right\}$s is considered. It can be observed from Fig.~\ref{Passive_receiver_comparison} that the proposed scheme still retains the superiority for all considered cases. Besides, we can see that the proposed scheme with the theoretical $n_u^*$ obtained from \eqref{final_tu_passive} is approximately matched to that with the numerical $n_u^*$ obtained from \eqref{problem_formulation_based_on_mSINAR_passive} and the ideal scheme, showing the accuracy of the derived $n_u^*$ and the effectiveness of mSINAR for the passive receiver. However, for some cases (such as $T_s=1$s and $L=3$), the BER curve with the theoretical $n_u^*$ is perfectly matched to that with the numerical $n_u^*$; while for other cases (such as $T_s=1$s and $L=10$), the gap between the BER curves with the numerical/theoretical $n_u^*$ is relatively large. This can be attributed to the fact that the received signal is discrete for the passive receiver with limited sampling times, and thereby, $n_u^*$ actually corresponds to a period of $\left[ {\left( {n_u^* - \frac{1}{2}} \right){t_s},\left( {n_u^* + \frac{1}{2}} \right){t_s}} \right]$ rather than an exact time similar to $t_u^*$, causing the mentioned fluctuation.


\section{Conclusion}\label{Conclusion}

For MCvD systems with an optimal detection interval, in this letter, we proposed to reuse the discarded duration outside the above detection interval to further eliminate the ISI. Besides, we formulated an objective function related to the discarded time to optimize the BER performance, where mSINAR is used to reduce the computational complexity. Moreover, an optimally reusable duration within the discarded time was derived in closed form. Monte Carlo simulations were performed to study the BER performance of the proposed ISI mitigation scheme. It is shown that compared with the benchmarks, the proposed scheme can achieve significant BER performance gain, especially in the case of severe ISI.

\begin{appendices}

\section{}

According to [11, eq.~25], we can convert mSINAR to modified signal-to-interference (mSID) to simplify the optimization procedure, i.e., \eqref{problem_formulation_based_on_mSINAR} can be written~as
\begin{align}\label{problem_formulation_based_on_mSID}
\hspace{-0.1cm}{\tilde t}_u^*\hspace{-0.05cm} \approx\hspace{-0.05cm}\left\{ \begin{array}{l}
\vspace{-0.2cm}
\hspace{-0.1cm}\mathop {\arg \max }\limits_{0 \le {t_u} < t_1^*} \Bigg\{\sum\limits_{k = 0}^L {{{\left( { - 1} \right)}^{\left| {k \cap 0} \right|}}\left( {F_{\left( {0,t_u} \right)}^k \hspace{-0.05cm} -\hspace{-0.05cm}  F_{\left( {t_1^*,t_2^*} \right)}^k} \right)}
\\
\vspace{-0.1cm}
\\
\hspace{-0.05cm} -\hspace{-0.05cm} \sqrt {\frac{2}{{Q}}} \sum\limits_{k = 0}^L {\sqrt {F_{\left( {t_1^*,t_2^*} \right)}^k\left( {1 \hspace{-0.05cm} -\hspace{-0.05cm}  F_{\left( {t_1^*,t_2^*} \right)}^k} \right)} }\Bigg\} ,~{\textrm{if}}~0 \hspace{-0.05cm}<\hspace{-0.05cm} {Q} \hspace{-0.05cm}< \hspace{-0.05cm}\hat{Q}
\\
\vspace{-0.1cm}
\\
\hspace{-0.1cm}\mathop {\arg \max }\limits_{0 \le {t_u} < t_1^*}\Bigg\{ \sum\limits_{k = 0}^L {{{\left( { - 1} \right)}^{\left| {k \cap 0} \right| }}\left( {F_{\left( {0,t_u} \right)}^k \hspace{-0.05cm} -\hspace{-0.05cm}  F_{\left( {t_1^*,t_2^*} \right)}^k} \right)}
\\
\vspace{-0.1cm}
\\
\hspace{-0.05cm} - \hspace{-0.05cm} \sqrt {\frac{2}{\hat {Q}}} \sum\limits_{k = 0}^L {\sqrt {F_{\left( {t_1^*,t_2^*} \right)}^k\left( {1 \hspace{-0.05cm} -\hspace{-0.05cm}  F_{\left( {t_1^*,t_2^*} \right)}^k} \right)} } \Bigg\},~{\textrm{if}}~{Q} \hspace{-0.05cm} \ge\hspace{-0.05cm} \hat{Q} \\
 \end{array} \right.,
\end{align}
where ${\left| {k \cap 0} \right|}$ is the cardinality of a set ${k \cap \left\{0\right\}}$. Different from \eqref{problem_formulation_based_on_mSINAR}, the impact from noise in $\left[ {0,{t_u}} \right]$ has been neglected in \eqref{problem_formulation_based_on_mSID}. This is because for ${0 \le {t_u} < t_1^* \ll {T_s}}$, we can have $F_{\left( {0,{t_u}} \right)}^i\big( {1 - F_{\left( {0,{t_u}} \right)}^i} \big) \ll F_{\left( {t_1^*,t_2^*} \right)}^i\big( {1 - F_{\left( {t_1^*,t_2^*} \right)}^i} \big)$. Besides, we assume that $\left[ {t_1^*,t_2^*} \right]$ is known for the target MCvD system. Based on this assumption, \eqref{problem_formulation_based_on_mSID} can be further simplified as
\begin{align}\label{simpified_problem_formulation_based_on_mSID}
{\tilde t}_u^* \approx \mathop {\arg \max }\limits_{0 \le {t_u} < {\bar t_1^*}} \int_{\rm{0}}^{t_u} {\left[ {\sum\limits_{k = 1}^L {h\left( {t + k{T_s}} \right)}  - h\left( t \right)} \right]dt},
\end{align}
where $\bar t_1^*$ is the value to which $t_1^*$ eventually converges. It is obvious that \eqref{problem_formulation_based_on_mSID} is dependent on $Q$ and according to the investigation in \cite{Xuan_previous_work}, $t_1^*$ also depends on $Q$. Thereby, we update $t_1^*$ as $\bar t_1^*$ to make \eqref{simpified_problem_formulation_based_on_mSID} hold true. Besides, we can see from \eqref{simpified_problem_formulation_based_on_mSID} that when $t \in \left[ {0,{\bar t_1^*}} \right)$ and $T_s>t_{\max}$, ${h\left( t \right)}$ is an increasing function, while ${\sum\limits_{k = 1}^L {h\left( {t + k{T_s}} \right)} }$ is a decreasing function, where ${t_{\max }}$ represents the peak time for the molecule concentration when an impulse of molecules is emitted at $t = 0$. This means that \eqref{simpified_problem_formulation_based_on_mSID} is equivalent to collecting all intervals satisfying $\sum\limits_{k = 1}^L {h\left( {t + k{T_s}} \right)}  - h\left( t \right) \ge 0$, and hence, we have
\begin{align}\label{further_simpified_problem_formulation_based_on_mSID}
\sum\limits_{k = 1}^L {h\left( {{\tilde t}_u^* + k{T_s}} \right)}  - h\left( {{\tilde t}_u^*} \right) = 0.
\end{align}
We find \eqref{further_simpified_problem_formulation_based_on_mSID} to be almost identical to [11, eq.~24]. Therefore, we just provide the final solution for \eqref{further_simpified_problem_formulation_based_on_mSID}, as shown in \eqref{final_tu_AB}.
For the specific solution process, please refer to [11, eq. 24].

\section{}
Similar to \eqref{problem_formulation_based_on_mSID}, we also convert mSIANR to mSID to simplify the calculation of \eqref{problem_formulation_based_on_mSINAR_passive}, which can be rewritten~as
\begin{align}\label{problem_formulation_based_on_mSID_passive}
{\tilde n}_u^* \approx \mathop {\arg \max }\limits_{0 \le n_u < {\bar n_1^*}} \left( {\sum\limits_{k = 1}^L {\sum\limits_{n = 0}^{n_u} {{p_{n,k}}} }  - \sum\limits_{n = 0}^{n_u} {{p_{n,0}}} } \right),
\end{align}
where $\bar n_1^*$ is the value of the final convergence of $n_1^*$. In \eqref{problem_formulation_based_on_mSID_passive}, the variance of the noise in $\left[ {0,n_u} \right]$ has been neglected, since it is much smaller than the variance of the noise in $\left[ {n_1,n_2} \right]^*$ and $\left[ {n_1,n_2} \right]^*$ is assumed to be known in this paper. Following \eqref{simpified_problem_formulation_based_on_mSID}-\eqref{further_simpified_problem_formulation_based_on_mSID}, we also have that if
\begin{align}\label{simplified_problem_formulation_based_on_mSID_passive}
{p_{{\tilde n}_u^*,0}} = \sum\limits_{k = 1}^L {{p_{{\tilde n}_u^*,k}}}
\end{align}
can be solved, we can obtain ${\tilde n}_u^*$. According to [11, eq.~37], the solution of $n_u^*$ can be expressed~as \eqref{final_tu_passive}. Next, we need to~ensure $\sum\limits_{k = 1}^L {{p_{n_u^*,k}}}  > {p_{n_u^*,0}}$ rather than ${p_{n_u^*,0}} > \sum\limits_{k = 1}^L {{p_{n_u^*,k}}}$,~thereby the floor function rather than ceiling function is used in~\eqref{final_tu_passive}.

\end{appendices}

\bibliographystyle{IEEEtran}
\bibliography{IEEEabrv, Final_manuscript_CL_paper}

\end{document}